# A Multi-scale Video Denoising Algorithm for Raw Image

Bin Ma, Yueli Hu*, Xianxian Lv and Kai Li

School of Mechatronic Engineering and Automation, Shanghai University, Shanghai 200444, China
`huyueli@shu.edu.cn`

**Abstract.** Video denoising for raw image has always been the difficulty of camera image processing. On the one hand, image denoising performance largely determines the image quality, moreover denoising effect in raw image will affect the accuracy of the following operations of ISP processing flow. On the other hand, compared with image, video have motion information in time sequence, thus motion estimation which is complex and computationally expensive is needed in video denoising. In view of the above problems, this paper proposes a video denoising algorithm for raw image, performing multiple cascading processing stages on raw-RGB image based on convolutional neural network, and carries out implicit motion estimation in the network. The denoising performance is far superior to that of traditional algorithms with minimal computation and bandwidth, and has computational advantages compared with most deep learning algorithms.

## 1 Introduction

Noise may be produced in all parts of the imaging system, such as sensor signal acquisition , signal transmission and subsequent signal processing. The imaging quality of digital imaging equipment is easily disturbed by noise. Although the image sensor has made great progress to reduce noise in recent years, denoising operation is still an inevitable part in video processing. Compared with image denoising, more scene information can be used in video denoising, which is helpful to the denoising process. However, video denoising needs better time correlation, which makes the requirements of denoising algorithm higher. The research on video denoising algorithm is of great significance to improve the visual quality of video, especially the imaging performance under challenging conditions(low illumination, small sensors, etc.), and provide more high-quality initial data for subsequent image analysis.

At present, most denoising methods are aimed at RGB image. However, raw image directly obtained by the image sensor contains abundant original information, and its noise model is relatively simple. After ISP, although the image has been denoised in some degree, the noise model of the remaining noise has become more complex and difficult to deal with. So it is more appropriate to carry out denoising process in the front of ISP in order to improve the effect of denoising. Therefore, this paper proposes an end-to-end multi-stage recursive video denoising algorithm performing multiple cascading processing stages for raw image: First, the continuous multi-frame raw images of a video is fed into the fusion stage as input, and recursively combined to reduce the noise; Second, the fused image is sent into the denoising stage to further remove remaining noise; Finally, the denoised image is sent into the refinement stage to recover the high-frequency details and improve the image quality. In addition, the algorithm is a three-scale recursive network. Different scale networks share the same parameters, that makes the model learning ability more stable and denoising performance better. Experiments on raw image show that it can surpass more complex algorithms with less computational cost.

## 2  Related works

### 2.1  Single frame denoising

In 2016, Zhang et al. proposed a convolution network combining batch normalization and residual structure called DnCNN[1] for image denoising, which is one of the earliest articles using deep learning model for denoising. The image with noise is processed by a series of feed forward convolution network, and the output of the network uses residual learning to generate a residual map containing only noise. Experiments show that residual learning and batch normalization complement each other, which improves the training stability and denoising effect. It can even use a single model to deal with the problems in the fields of Gaussian denoising, super-resolution and JPEG decoding, and have good generalization ability of restored image. However, the rapidity and convergence of the whole algorithm are not prominent enough.

In 2018, Zhang et al. improved the DnCNN and proposed a fast and flexible deep convolution denoising network called FFDNet[2]. Taking the noise estimation as the input of the network can deal with more complex noise and weigh the suppression of noise and the maintenance of details. The main feature of the network is that it can use a single network to deal with different levels of noise and more complex real scenes, moreover improve the speed of the algorithm while maintaining the denoising performance.

REDNet[3] was the first to use self-coding structure for image recovery, with encoding composed of convolution layer and decoding composed of deconvolution layer. This method extracts the main features of the input image through encoder, restores the semantic details of the image through decoder, and then directly predicts the noise. Finally, the clean image can be obtained by subtracting from the image containing noise. By using skip layer connection to symmetrically connect the encoding and decoding, the training convergence speed is faster and the local optimal value of higher quality can be obtained .

Liu et al.[4] proposed Wide Inference Network(WIN) based on convolutional neural network to study the influence of the"depth" and "width" of network on noise reduction effect. He came to the conclusion that wider network (with more channels and larger convolutional kernels) have a larger receiving area and more neurons in the convolutional layer, which can learn pixel distribution characteristics more effectively, and will make the noise reduction effect better.

Divakar et al.[5] proposed an image blind denoising model using the generated adversarial network(GAN), which does not require prior information. The input is a single noisy image and the output is a clear image. Through joint optimization of the denoising network and the discriminant network, the processing effect of the denoising network was improved. In view of the difficulty in obtaining paired training samples, Chen et al.[6] used GAN to model the noise information extracted from the real noise image, and combined the noise block randomly generated by the generator with the original clear image to synthesize a realistic noise image. The real image dataset has been expanded, and the problem of insufficient paired training samples has been alleviated to a certain extent.

### 2.2 Video denoising

Non-local self-similar denoising algorithm uses the similarity of each pixel in the image to denoise. Typical algorithms include non-local Means (NLM) [7], Block-Matching and 3D Filtering (BM3D)[8], etc. It searches for similar areas in the image in units of image blocks, and then averages these areas, which can better remove the gaussian noise in the image. Different from bilinear filtering and median filtering which commonly use image local information filtering, it uses the whole image to denoise. This algorithm is fit for image with regular and repetitive detail features. But it usually requires manual selection of parameters and requires a lot of time to optimize. Moreover classical BM3D algorithms can also be extended to video image sequences, namely VBM3D[9] and VBM4D[10] algorithms.

In recent years, deep learning networks with strong learning ability and flexible architecture have set off a research boom in the field of video denoising.

DVDNet[11] is an explicit two-stage denoising network. It first performs convolution denoising on the current frame and neighboring frame in the spatial domain, then aligns the neighboring frame with the current frame through optical flow network, and finally performs time domain denoising on the aligned image. However, this result will very much depend on the accuracy of the optical flow estimation.

Therefore, the author proposed FastDVDNet[12], which abandoned optical flow alignment and replaced it with another two-stage network.First, the long-term image sequence is divided into several shorter and overlapping image sequences, and this sequence is sequentially passed through the weight-sharing noise reduction network to complete the preliminary noise reduction.Then, the output is further denoised through another network whose structure is shared but whose weights are not shared. The two levels make full use of the redundant information in short and long time.

### 2.3 Raw image denoising

At present, most image denoising methods are for RGB image, However, image has lost a lot of original information from the raw image to the RGB image, so the more ideal processing method should be to complete the noise reduction in the raw image.

Megvii[13] proposed a mode unification and augmentation method for raw image in NTIRE 2019. Raw image is obtained directly from the camera sensor, with Bayer mode four channels. Due to hardware reasons, there are four sorts of the four channels: GRBG, BGGR, GBGR, RGGB. This paper proposes BayerUnify, which uses crop and padding to unify the data in Bayer format entering the network . Due to the lack of dataset of raw image, only data enhancement can be used to expand the data. However, raw data is different from RGB format, data enhancement operations such as arbitrary cropping and flipping of the image will cause the pixel channel space arrangement to be disordered, and the real image differ greatly. This paper proposes BayerAug , which combines flipping and cropping to perform data enhancement while maintaining the Bayer pattern of the image.

Google[14] published an article on CVPR 2019, which was mainly aimed at solving the problem of dataset construction. By analyzing each flow in ISP, the author tries to find the inverse of each transformation. According to reverse the flow of ISP, the sRGB image is transformed into raw image, and then used for the training of convolutional neural network, so as to achieve the process of noise reduction.

In recent years, a series of improved methods combining residual structure, multi-scale feature fusion and transfer learning have emerged continuously on the basis of CNN network. When studying the task of image denoise, it is necessary to select the suitable network according to the existing problems of the current image denoising algorithm and the problems to be solved.

## 3     Preliminaries

### 3.1    Noise model

Noise is an interference signal superimposed on the original image. The goal of the denoising algorithm is to estimate a clean video from the observed noise data. The observation model is defined as follow:

$$z_t(x) = y_t(x) + \eta_t(x) \tag{1}$$

Where $z_t(x)$ is the observed raw noise video, $y_t(x)$ is the noise-free data to be estimated, $\eta_t(x) \sim N(0, \sigma_t^2(y_t))$ is the noise observation, where the noise variance is:

$$\sigma_t^2(y_t) = a_t y_t + b_t \tag{2}$$

$a_t, b_t$ are parameters for signal-dependent (shot) and signal-independent (read) noise obtained from the noise calibration of a given camera and ISO[15], and the parameters are different for different ISO.

### 3.2    Preproccess

As this algorithm directly acts on raw image which directly obtained from the camera sensor and obeyed 4-channel Bayer pattern, instead of directly processing the raw image in CFA structure, raw image should be preprocessed and converted into RGGB image first. There are four different Bayer patterns due to hardware: GRBG, BGGR, GBGR and RGGB. The dataset used in this paper is 1-channel raw data obeyed Bayer pattern GBRG with unknown noise . In order to ensure the Bayer pattern of the real image, this paper refers to the method proposed in reference[13] to convert it into an RGGB 4-channel image by flipping and down-sampling on each frame. Thus, a complete 4-channel image sequence is obtained, which is represented by $\{I_n\}_{n=1}^{N}$ ,that is $I_n = (I_n^R, I_n^{G1}, I_n^{G2}, I_n^B)$ ,and each new frame is half the width and height of the original data.

### 3.3    Color transformation

The color transformation $T_c$ is a decorrelation transformation.The 4-channel image sequence is transformed into a channel uncorrelation space, and RG1G2B is converted into YUVW[16].It is realized as a point-wise convolution whose kernel constraint is an orthonormal matrix $M \in R^{C \times C}$ . The inverse matrix $T_c^{-1}$ generates an inverse transformation, which transfers the image from YUVW back to the RG1G2B space.

$$L_c = \|M \cdot M' - I_C\|_F \tag{3}$$

Where, $I_C$ is the identity matrix of rank C, $\|\cdot\|_F$ is the Frobenius norm, and the matrix $M \in R^{1 \times 1 \times C \times C}$ is the kernel of the convolution layer (point convolution), which is initialized as:

$$M = \begin{bmatrix} 0.5 & 0.5 & 0.5 & 0.5 \\ -0.5 & 0.5 & 0.5 & -0.5 \\ 0.65 & 0.2784 & -0.2784 & -0.65 \\ -0.2784 & 0.65 & -0.65 & 0.2784 \end{bmatrix} = \begin{bmatrix} Y \\ U \\ V \\ W \end{bmatrix} \tag{4}$$

## 4    Proposed method

At present , the research on denoising of the raw image is limited, and the existing models are relatively large and complex, and require high computing power. In order to achieve a denoising performance far beyond traditional algorithms with very small computing power, this paper proposes an end-to-end video denoising algorithm performing multiple cascading processing stages for raw image. Figure 1 is the algorithm processing structure.

Based on the convolutional neural network, multi-scale cascading processing operations of raw-RGB image are carried out, which consists of three parts, namely: fusion stage, denoising stage and refinement stage. The fusion stage initially removes noise through the fusion of all past video frames; And the refinement stage is responsible for restoring the detailed information lost in the denoised image. At the same time, this algorithm proposes to add a recursive connection between three different scales of network and share the weights among different scales, so as to reduce network parameters and improve model learning ability.

### 4.1 Multi-scale

Inspired by the literature[17], this paper designs a three-scale network structure. In this paper, the input image is down-sampled into different scales through stride convolution (stride is 2,4,8 respectively), and placed in different scale layers for processing. And at the same time the idea of residual is introduced. In the stage of denoising stage, when there are multiple scales, Then, the fusion operation is carried out to make better use of the feature information between different scales. By the idea of coarse-to-fine, finer denoising results can be obtained.

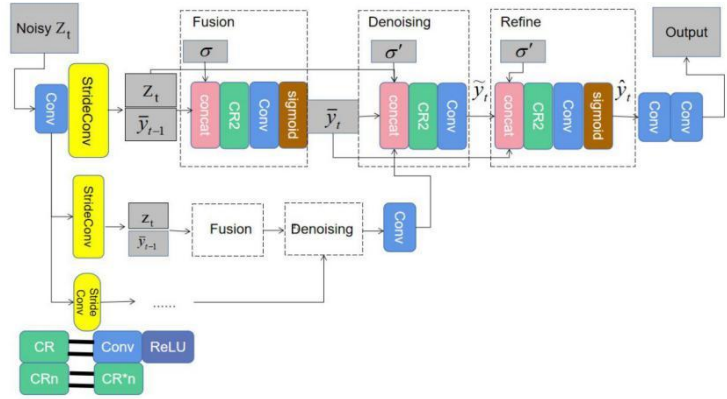

**Fig.1.** Algorithm network structure.

Conv means convolution module, StrideConv means strided convolution module , CR2 is an user-defined module consisting of basic modules.

### 4.2 Fusion module

Since the object of this paper is a video which has more available information in upper and lower frames than a single frame image, the goal of the fusion module is to use the inherent time correlation of the video to minimize the noise in the image.
The fusion stage is defined as follows, and the output is the weight of the fusion stage predicted by the network:

$$\gamma_t(x) = Fusion(|z_t - \bar{y}_{t-1}|, \hat{\sigma}_t^2) \tag{5}$$

Where $z_t$ is the current noise input frame after the color transformation, $\bar{y}_{t-1}$ is the previous fused frame, $\hat{\sigma}_t^2 = \sigma_t^2(z_t)$ is the noise variance of the input frame, and $|\cdot|$ denotes absolute value. For the first frame t=0, since there is no previous frame, the initial condition is $\bar{y}_{t-1}(x) \equiv z_t(x)$.

The implementation is shown in Figure 2 below. Two convolutional layers(3 × 3 kernels) are followed by a ReLU activation layer and a convolutional output layer. The output layer is followed by Sigmoid activation to ensure saliency.
Fusion is defined as the following recursive function:

$$\bar{y}_t(x) = \bar{y}_{t-1}(x)\bar{\gamma}_{t-1}(x) + z_t(x)\gamma_t(x) \tag{6}$$

Where $\bar{y}_t(x)$ is the current output frame after fusion, $z_t$ is the current transformed noisy input frame, $\bar{y}_{t-1}(x)$ is previous fused frame, $\gamma \in R^{H/2 \times W/2 \times 1}$ is the fusion network weight obtained by (5), which satisfied $\bar{\gamma}_{t-1}(x) + \gamma_t(x) = 1$. Since the number of channels of weight is 1, the fusion of all 4C input channels is realized through element-wise broadcasting.

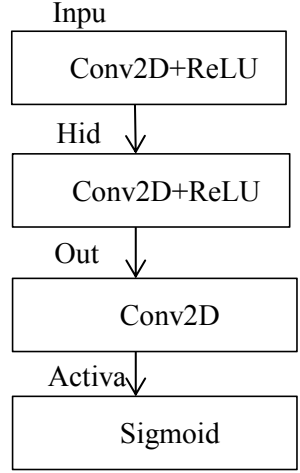

**Fig. 2.** Implementation of the fusion stage architecture.

### 4.3 Denoising module

Noise in image is reduced by fusion module but not completely, so this paper uses a denoising module to further remove remaining noise in fused image. The denoising network is defined as:

$$\tilde{y}_t = Denoising(\bar{y}_t, z_t, \bar{\sigma}_t^2) \tag{7}$$

Where $\tilde{y}_t$ is the denoised image; The input also includes the current noise frame $z_t$, so that the network has the opportunity to extract valuable information from the noise input; $\bar{\sigma}_t^2$ is the noise variance of the fused image $\bar{y}_t$, which depends on the signal-dependent variance at current frame and the noise variance of all previous frames fused, so the recursive formula for noise variance of the fused image is:

$$\bar{\sigma}_t^2 \equiv \sigma_t^2(\bar{y}_t) = \gamma_t^2 \sigma_t^2(z_t) + \bar{\gamma}_{t-1}^2 \sigma_{t-1}^2(\bar{y}_{t-1}) \tag{8}$$

The denoising stage, which is similar to the fusion stage, is realized by connecting image and variance as input. This stage still uses two convolution layers (3 × 3 kernels) followed by ReLU activation layer and one convolution output layer, but the activation layer is not applied after the output layer.

### 4.4 Refinement module

The image will inevitably lose some details after denoising, therefore this paper constructs a refinement module, which combines the fused image $\bar{y}_t$ (detailed but still noisy) with denoised image $\tilde{y}_t$ (noise-free but oversmoothed). It identifies high-frequency information from the fused image to restore some details. Therefore, the refinement is defined as:

$$\hat{y}_t(x) = \bar{y}_t(x)\overline{\omega}_t(x) + \tilde{y}_t(x)\omega_t(x) \tag{9}$$

Where $\omega \in R^{H/2 \times W/2 \times 1}$ is the weight of the refinement network, satisfying the relationship $w_t(x) + \overline{w}_t(x) = 1$, which can be obtained from the refinement network prediction:

$$\omega_t(x) = \text{Refine}(\tilde{y}_t, \bar{y}_t, \bar{\sigma}_t^2) \tag{10}$$

The implementation is to take the connection of the fused image, denoised image and noise variance as input. And its structure is similar to the fusion network, and the output layer is also connected to Sigmoid activation.

## 5 Experiments

### 5.1 Dataset and preprocessing

This paper uses the dataset CRVD proposed in[15], which is the first dynamic video dataset with noise removal pairs. It creates motion over controllable objects such as toys, and each static moment is captured multiple times by the SONY IMX385 sensor to generate clean video frames. It contains 11 indoor scene videos, and all videos have five different ISO levels ranging from 1600 to 25600, which is a total of 3850 frames. In order to train the model, the dataset needs to provide noise-free image as Ground Truth (GT) . There are two main methods to obtain GT : one is to use a low ISO long exposure image as the GT , and the other is to merge multiple high ISO short exposure image as GT . The second method is adopted in this paper. For each static moment, 10 noise frames are captured continuously and averaged.

The obtained image is regarded as an average noise frame, but there is still a slight noise, so BM3D [8] is further applied to get a completely clean ground truth. This paper uses scenes 1-6 for training and 7-11 for objective validation. CRVD also contains another 10 outdoor noisy videos without GT as testset to subjectively evaluate the visual quality of denoising.

This paper randomly selects n= 25 images from 70 frames in each ISO of each scene as an original training sequence. Then they are cropped into a size of 128×128 at random spatio-temporal positions, meanwhile each image was randomly cut to s=16, forming 16 training sequences. After preprocessing the data in accordance with section 3.2, a 4-channel image sequence is obtained, which is represented by $\{I_n\}_{n=1}^{N}$, that is $I_n = (I_n^R, I_n^{G1}, I_n^{G2}, I_n^B)$.

Each new frame is half the width and height of the original data, so the noise data input of convolutional network after preprocessing is $z_t \in R^{16*25*4*64*64}$.

## 5.2 Experimental process

The environment used in the experiment is shown in the following Table 1:

**Table 1.** Experimental environment on the server.

| Operating system | Ubuntu 16.04 |
| --- | --- |
| GPU | GeForce GTX 2080 Ti |
| CUDA | 10.0 |
| Experimental environment | python3.8 |
| Dependency Library | Python numpy opencv Pytorch skimage opencv tensorboard etc |

The loss function of the network is defined as:

$$L = L_r + L_c \tag{11}$$

Where $L$ is the total loss, including $L_c$ is the constraint term for the reversibility of the color transformation, and $L_r$ is the loss term that measures the average difference between the predicted value of a training sequence and the true value. Expression of $L_r$ is:

$$L_r = \frac{1}{n}\sum_{t=1}^{n}\|\hat{y}_t - y_t\|_1 \quad (\|\cdot\|_1 \text{ is the mean L1 norm}) \tag{12}$$

This article uses Adam optimizer with batch size 8 and initial learning rate 1e-4 to train the model. This paper uses a segmented learning rate constant decay, which sequentially decreased by a factor of 10 when 70% and 90% of the iterations are completed. A total of 3000 epochs iterations are performed in the training, and validation is performed every 25 epochs.

## 5.3 Experimental results

Figure 3 shows the trend of total network loss as the number of iterations increases when using GPU for training epochs=3000. The abscissa is the number of iterations, and the ordinate is the value of the total loss.

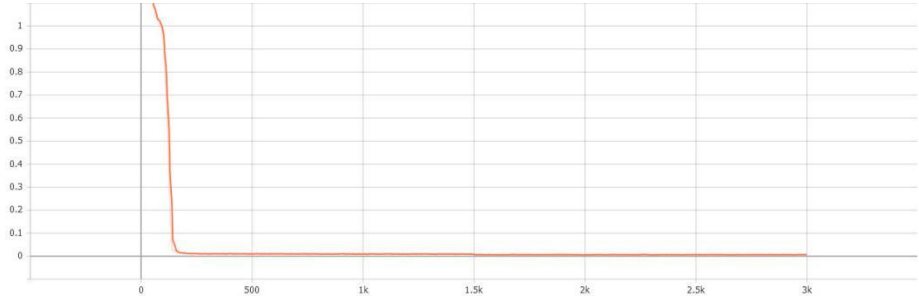

**Fig. 3.** Variation trend of network total loss.

It shows that the loss term $L$ declines rapidly after a brief slowdown at epochs = 80, eventually converged to around 7e-3, indicating that the model is reasonably designed and can converge quickly and effectively. The loss $L$ consists of two parts $L_c$ and $L_r$. At the later stage of training, $L_c$ is converged to 0 indicating that the color relation of the picture is strictly constrained, and the loss $L$ is all caused by $L_r$. Since $L_r$ measures the gap between the model prediction and the truth value, the loss converged close to zero indicates that the trained model has good performance and the output prediction is close to the true value. Figures 4 and 5 show the trend of PSNR and SSIM of denoising effect respectively for training epochs=3000 using GPU.

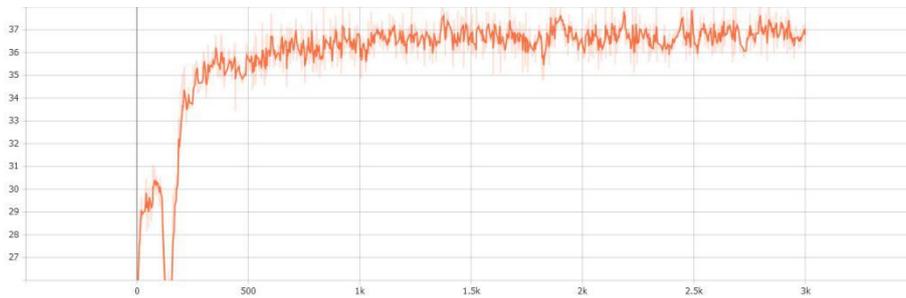

**Fig. 4.** Variation trend of PSNR.

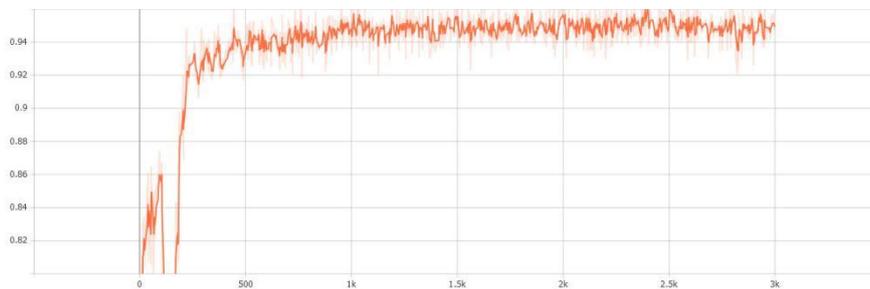

**Fig. 5.** Variation trend of SSIM.

The abscissa is the number of iterations, and the ordinate is respectively the value of PSNR and SSIM obtained by verifying the model on the validation set under ISO=25600.

It can be seen that as the training going on, except for a fluctuation when epochs=100, the PSNR and SSIM keep rising till about 37.5 and 0.96 respectively, indicating that the model has a good denoising performance.

Table 2 shows the mean values of PSNR and SSIM taking scenes 7-11 as validation set in different ISO levels.

In order to evaluate the visual denoising performance of the model, this paper randomly selects low-illumination outdoor scenes(scene=1,2,10) to test the model training for epochs = 3000, and obtain the following visual effect images. The left side are the original noisy images, and the right side are the output denoised image by the model. In order to feel the noise and noise removal effect more intuitively, this paper chooses the noisy image with

ISO=25600 for denoising. It can be seen that the noise of image is significantly reduced. Even in the dark condition, the visual denoising performance is excellent and the image becomes smoother after denoising.

**Table 2.** Mean values of PSNR and SSIM of model respectively testing on validation set at different ISO levels.

| ISO | Noisy | | Model (Epochs=1500) | | Model (Epochs=3000) | |
|---|---|---|---|---|---|---|
| | PSNR | SSIM | PSNR | SSIM | PSNR | SSIM |
| 1600 | 38.922 | 0.951 | 44.361 | 0.990 | 44.639 | 0.990 |
| 3200 | 35.808 | 0.917 | 42.398 | 0.986 | 43.646 | 0.989 |
| 6400 | 32.448 | 0.878 | 40.694 | 0.980 | 41.544 | 0.981 |
| 12800 | 28.174 | 0.766 | 38.228 | 0.974 | 38.065 | 0.973 |
| 25600 | 26.510 | 0.670 | 37.713 | 0.960 | 37.796 | 0.963 |
| Average | 32.3724 | 0.8364 | 40.6788 | 0.978 | 41.138 | 0.9792 |

Table 3 compares the parameters of the algorithm in this paper and the DnCNN[1] algorithm. As can be seen from the table, and number of parameters and floating point operations of the algorithm in this paper are far smaller than DnCNN algorithm. The parameters of this algorithm are 0.32M, and the floating point Multiplier and Accumulation (Mac) operations of a sequence are amount of 1.42 GMac. Therefore, the time and calculation power consumed by the algorithm in this paper are smaller than that of DnCNN algorithm in testing dataset. Considering the objective and subjective indicators of denoising performance, the algorithm proposed in this paper is a better one.

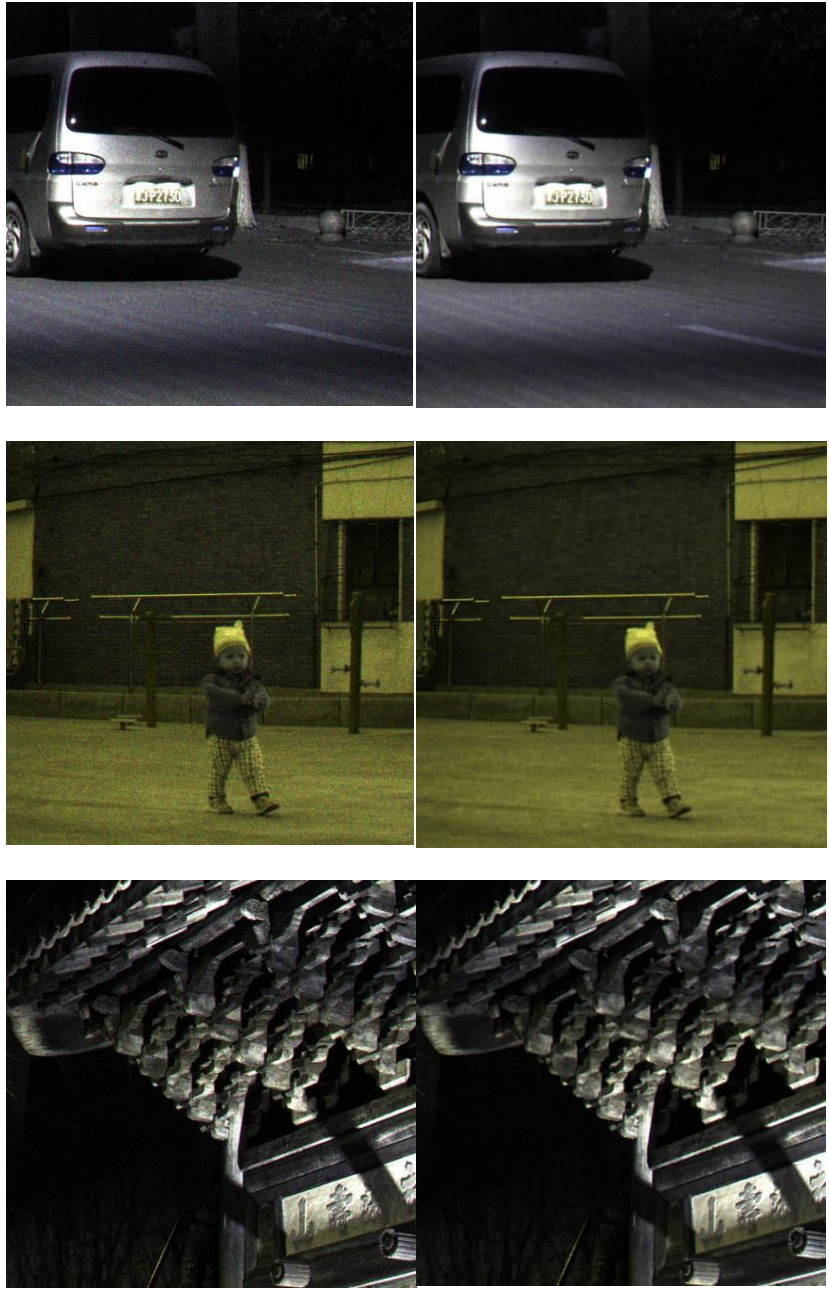

**Fig. 6.** Visual denoising performance of the model on the outdoor scene=1, 2 and 10.

**Table 3.** The parameters of the algorithm compared between this paper and DnCNN.

| Method | Parameters | GMac |
|---|---|---|
| DnCNN | 0.56M | 18.25 |
| Ours | 0.32M | 1.42 |

## 6      Conclusion

In order to improve the visual imaging quality of video, especially to improve the imaging performance of the camera in low illumination environment, this paper proposes an end-to-end multi-scale recursive video denoising algorithm for the raw image through multiple cascading processing stages, namely fusion stage, denoising stage, and refinement stage.The frames are preliminarily reduced noise through multi-frame fusion by the fusion stage. Through multi-scale recursion and sharing parameters between different network scales make the learning ability of the model more stable. Experimental results show that both quantitative and qualitative indicators show good denoising performance on indoor and outdoor videos. The algorithm in this paper can surpass more complex algorithms with less computing power.